\documentclass[fleqn]{article}
\usepackage{bbding}
\usepackage{latexsym}
\usepackage[latin1]{inputenc}
\usepackage{graphicx}

\newcommand{\T}{{\rm Tr}}

\newcommand{\cH}{{\cal H}}

\newcommand{\bA}{{\mathbf A}}
\newcommand{\bG}{{\mathbf G}}
\newcommand{\rd}{{\mathrm d}}
\newcommand{\0}{{\mathbf 0}}
\newcommand{\1}{{\mathbf 1}}

\def\bbbc{{\mathchoice {\setbox0=\hbox{$\displaystyle
\rm C$}\hbox{\hbox
to0pt{\kern0.4\wd0\vrule height0.9\ht0\hss}\box0}}
{\setbox0=\hbox{$\textstyle\rm C$}\hbox{\hbox
to0pt{\kern0.4\wd0\vrule height0.9\ht0\hss}\box0}}
{\setbox0=\hbox{$\scriptstyle\rm C$}\hbox{\hbox
to0pt{\kern0.4\wd0\vrule height0.9\ht0\hss}\box0}}
{\setbox0=\hbox{$\scriptscriptstyle\rm C$}\hbox{\hbox
to0pt{\kern0.4\wd0\vrule height0.9\ht0\hss}\box0}}}}

\begin{document}
  \title{Transition Probability (Fidelity) and its Relatives}
  \author{Armin Uhlmann}
  \date{University of Leipzig,\\ Institute for Theoretical Physics}
  \maketitle

\begin{abstract}

Transition Probability (fidelity) for pairs of density operators
can be defined as a ``functor'' in the hierarchy of
``all'' quantum systems and also within any quantum system.
The Introduction of ``amplitudes'' for density operators
allows for a more intuitive treatment of these quantities,
also pointing to a natural parallel transport. The latter
is governed by a remarkable gauge theory with strong relations
to the Riemann-Bures metric.

\end{abstract}
\section{Introduction}
The topic of the paper concerns transition probability and
fidelity for general (i.~e. mixed) states and some of its descendants.
It belongs, metaphorically spoken, to the ``skeleton'' or
to the ``grammar'' of Quantum Physics in which dynamics
does not play a significant role. The needs of Quantum
Information Theory have considerably pushed forward this abstract
part of Quantum Theory.

Transition probability between pure states is one
of the most important notions in Quantum Physics. It is basic
within the probability interpretation as initiated by
M.~Born and pushed into a general form by P.~A.~M.~Dirac,
J.~von Neumann, G.~Birkhoff and many others.

Transition probabilities for pure states, expressed by vectors
of a Hilbert space, are a standard text book issue:
Let $\cH$ be a Hilbert space and $\langle.,.\rangle$ its
scalar product.  Let $|\psi_1\rangle$ and $|\psi_2\rangle$
be two of its unit vectors. Then
\begin{equation} \label{intro1}
\Pr(\pi_1 , \pi_2) := \T \, \pi_1 \pi_2 =
| \langle \psi_1 , \psi_2 \rangle |^2
\end{equation}
is their transition probability.
$\pi_j = |\psi_j \rangle\langle \psi_j|$ denote the density operators
of the relevant pure states.
This primary meaning of (\ref{intro1}) is as following:
Let us assume the quantum system is in its pure state $\pi_1$.
Asking by a measurement whether the system is in state $\pi_2$
or not, there are two cases:
Either we obtain ``YES'' or ``NO''. If the answer is ``Yes'',
the state $\pi_2$ has been prepared. if the answer is ``NO''
the state with the vector $(\1 - \pi_2)|\psi_1\rangle$
has been prepared. It is
by mere chance which case takes place in an individual
measurement. The {\em probability} to get the
answer ``YES'' is equal to the transition probability
(\ref{intro1}), showing up approximately as the success
rate for a large number of cases.
\medskip

To circumvent heavy mathematical technics, see
\cite{AU00}\footnote{It
is possible to work within the category
of von Neumann or of unital $C^*$algebras.},
we restrict ourselves to ``full'' quantum systems
based on finite dimensional Hilbert spaces $\cH$.
States are represented by density operators.
Channels are completely positive and trace
preserving maps. We use the convention
\begin{displaymath}
\hbox{ fidelity } = \sqrt{ \hbox{ transition probability } } \; .
\end{displaymath}

\section{Transition probabilities for density operators}
What can be done if the system is in a mixed state with
density operator $\rho_1$ and we like to prepare another
mixed state, $\rho_2$, by a measurement? This task cannot
be performed within the system itself\footnote{Except
$\rho_2$ is pure.}. We have to leave the system based on
$\cH$ and have to go to larger systems in which one can
perform appropriate ``YES -- NO'' measurements as used
above to give to $\Pr(\rho_1, \rho_2)$ a clear meaning.

As it turns out, we do not have to consider arbitrary
large quantum systems for this task.
It suffices to work within $\cH \otimes \cH$.
Taking this for granted, we assume that $|\varphi_1\rangle$,
$|\varphi_2\rangle$ are purifying vectors for of
$\rho_1$, $\rho_2$ in $\cH \otimes \cH$ :
The partial trace of
\begin{equation} \label{intro2}
\pi'_j = |\varphi_j \rangle\langle \varphi_j| , \quad
|\varphi_j\rangle \in \cH \otimes \cH
\end{equation}
has to be $\rho_j$ for $j=1,2$. By the reasoning above
we obtain the probability
$|\langle\varphi_1, \varphi_2\rangle|^2$ to prepare the
state $\pi'_2$  from $\pi'_1$ by a suitable measurement.

This value is not uniquely attached to the pair $\rho_1$,
$\rho_2$. Generally, different purifications give different
values. However, within all these values there is a largest
one  and this is called the {\em transition probability}
from $\rho_1$ to $\rho_2$ and it will be denoted by
$\Pr(\rho_1, \rho_2)$.

In other words:  There are measurements in larger systems
preparing $\rho_2$ from $\rho_1$ with probability
$\Pr(\rho_1, \rho_2)$. But one cannot do it better.
Thus, \cite{Uh76a},
\begin{equation} \label{intro3}
\Pr(\rho_1, \rho_2) = \max_{{\rm all \, purifiations}}
| \langle \varphi_1, \varphi_2 \rangle |^2
\end{equation}
where the ``purification conditions''
\begin{equation} \label{intro4}
\T \, \rho_j A = \langle \varphi_j, (A \otimes \1) \varphi_j \rangle
 , \quad j = 1,2
\end{equation}
must be satisfied for all operators $A$ acting on $\cH$.

\subsection{Transition probability and channels}
With the increase of the quantity (\ref{intro1}),
the possibility to distinguish $\psi_2$ from $\psi_1$ becomes
more and more difficult. In the above definition of
$\Pr(\rho_1, \rho_2)$ we had to look for a pair of purifying
states the discrimination of which is as difficult as possible.
In this sense the problem (\ref{intro3}) is ``inverse'' to
that of state discrimination, \cite{J094}.

Now, sending two states through a quantum
channel, the possibility of their discrimination is diminishing.
This should imply a better chance to convert one of the images
into the other one and, therefore, should result in a larger
transition probability between the output states than between
input ones. This, indeed, is true. Let us make this more
transparent.

Cum grano salis we live in a ``quantum world''
consisting of an hierarchy of quantum systems. The physical
meaning of an individual system is highly determined by its
``place'' within other quantum systems. Here we are interested
in the corresponding state spaces and in the quantum channels
acting on or between them.

We consider functions (``functors''), $Q = Q(.,.)$,
attaching a real number to any pair of density operators on
any quantum system. With this in mind let us assume the
following two conditions:
\begin{itemize}
\item For pairs of pure states, $\pi_1, \pi_2$, we require
\begin{equation} \label{intro5}
Q(\pi_1, \pi_2) = \T \, \pi_1 \pi_2 \equiv \Pr(\pi_1, \pi_2)
\end{equation}
\item For all quantum channels $\Phi$ and all pairs of density
operators:
\begin{equation} \label{intro6}
Q(\rho_1, \rho_2) \leq Q(\Phi(\rho_1), \Phi(\rho_2))
\end{equation}
\end{itemize}
At first we convince ourselves that for all $Q$ satisfying
(\ref{intro5}) and (\ref{intro6}) we get
\begin{equation} \label{intro7}
\Pr(\rho_1, \rho_2) \leq Q((\rho_1, \rho_2)) \; .
\end{equation}

To see this we return to the purification procedure. While
$\rho_1$, $\rho_2$ are living on $\cH$, their purifications
are pure density operators $\pi_j$ on some $\cH \otimes \cH'$.
Then, abbreviating the partial trace over $\cH'$ by $\T'$,
it is $\T' \pi_j = \rho_j$.
In the case of a finite dimensional Hilbert space
the maximum in (\ref{intro4}) is already attained
in $\cH \otimes \cH$ by certain purifications $\pi_1$ and
$\pi_2$.  With them it follows
\begin{displaymath}
Q(\rho_1, \rho_2) = Q(\T' \pi_1, \T' \pi_2) \geq
Q(\pi_1, \pi_2) = \T \, \pi_1 \pi_2 = \Pr(\rho_1, \rho_2)
\end{displaymath}
and (\ref{intro7}) is established.

Does $\Pr(.,.)$ belong itself
to the set of functions obeying (\ref{intro5}) and
(\ref{intro6})? The answer is ``yes''. Indeed,
$\Pr(.,.)$ fulfills (\ref{intro6}) even for trace
preserving and just positive maps, \cite{AlUh83}.
By (\ref{intro7}) this guaranties
\begin{equation} \label{intro8}
\Pr(\rho_1, \rho_2) = \inf_{Q} Q(\rho_1, \rho_2)
\end{equation}
where $Q$ runs through all functions satisfying
(\ref{intro5}) and (\ref{intro6}).

While by (\ref{intro3}) the transition probability is
symmetric in its arguments, we did not require this
for a general $Q$ in (\ref{intro8}).

$\Pr(.,.)$ can be consistently extended to all
positive operators\footnote{of trace class if $\cH$ is
infinite dimensional} by
\begin{equation} \label{intro10}
\Pr(\lambda_1 \rho_1, \lambda_2 \rho_2) = (\lambda_1 \lambda_2)
\; \Pr(\rho_1, \rho_2) \; .
\end{equation}

Two examples follow for illustration. The first one reads
\begin{equation} \label{intro11}
(\T \, \omega_1)^{1-s} (\T \, \omega_1)^s \T \,
\omega_1^s \omega_2^{1-s} \geq \Pr(\omega_1, \omega_2)
\end{equation}
Because of (\ref{intro10}) it suffices to prove it for density
operators. However, one knows that
$Q := \T \, \rho_1^s \rho_2^{1-s}$
satisfies conditions (\ref{intro5}) and (\ref{intro6}).

Another example is
\begin{equation} \label{intro12}
4 \Pr(\omega_1, \omega_2) \leq (\T \, \omega_1 + \T \, \omega_2)^2
- \parallel \omega_1 - \omega_2 \parallel_1^2
\end{equation}
where $\parallel . \parallel_1$ denotes the trace norm.
(\ref{intro12}) is consistent with (\ref{intro10}). Its
right hand side respects (\ref{intro5}) and
(\ref{intro6}), proving the assertion.
Applied to density operators (\ref{intro12}) reduces to a know
inequality, see \cite{NieChubook} (exercise 9.21).

There are many other bounds, older \cite{Fu96} and newer ones,
\cite{MPHUZ09}, \cite{MNMFL09}.

\subsection{Explicit expressions}
The transition probability between two density operators
$\rho_1$ and $\rho_2$ of a quantum system can be evaluated
within that system. To do so, one needs explicit
expressions, \cite{Uh76a}. One can find them in
text books, for instance in \cite{NieChubook},
\cite{BZ06}. We present them in terms of fidelity:
\begin{equation} \label{expli1}
F(\rho_1, \rho_2)
= \T \, (\, \rho_1^{1/2} \rho_2 \rho_1^{1/2}\,)^{1/2}
= \T \, (\, \rho_2^{1/2} \rho_1 \rho_2^{1/2}\,)^{1/2}
\end{equation}
Remind that in the present paper we call ``fidelity''
the positive square root of transition probability.
Both, $F(.,.)$ and $\Pr(.,.)$ behave nicely with
respect to direct products:
\begin{equation} \label{expli2}
F(\rho_1 \otimes \rho_1' , \rho_2 \otimes \rho_2')
=
F(\rho_1 , \rho_2) \, F(\rho_1' , \rho_2')
\end{equation}
as follows directly from (\ref{expli1}).
\medskip

For later use we rewrite (\ref{expli1}) in a particular way,
suggested by the {\em geometrical mean} \cite{PW75, Ando79}.
defined For invertible positive operators
the latter is defined by
\begin{equation} \label{gem0}
\omega \# \rho = \rho^{-1/2}
(\rho^{1/2} \omega \rho^{-1/2})^{1/2} \rho^{-1/2}
\end{equation}
and it extends $\sqrt{\omega \rho}$ from
commuting pairs of positive operators to general ones,
see for instance \cite{BZ06} and \cite{CrUh} for more.
It can be seen from (\ref{gem0}) that $\omega \# \rho^{-1}$
is well defined by continuity for all pairs of positive operators,
wether invertible or not. To make it more transparent we
use the quasi-inverse $\omega^{[-1]}$ of $\omega$.
It enjoys the same eigenvectors as $\omega$, but all
{\em non-zero} eigenvalues are inverted. Now
\begin{equation} \label{ftp2a}
\omega \# \rho^{[-1]} = \rho^{[-1/2]} \,
(\rho^{1/2} \omega \rho^{1/2})^{1/2} \, \rho^{[-1/2]} \; .
\end{equation}
Finally, we rewrite (\ref{expli1}) in the following manner:
\begin{equation} \label{ftp2}
F(\rho_1, \rho_2)
= \T \, (\rho_2 \# \rho_1^{[-1]}) \rho_1
= \T \, (\rho_1 \# \rho_2^{[-1]}) \rho_2
\end{equation}

\section{Amplitudes}
Let $\rho$ be a density operator, not necessarily normalized.
We call {\em amplitude of} $\rho$ any operator $W$ which
satisfies
\begin{equation} \label{ampli1}
\rho = W W^{\dag} \; .
\end{equation}
The square root of a density operator is one of its amplitudes.

To be consistent we have to call ``amplitude'' of a pure state
$\pi = |\psi \rangle \langle \psi|$ any operator
$W = |\psi \rangle \langle \psi'|$ with unit vector $|\psi'\rangle$.
\medskip

If $W$ is an amplitude of $\rho$ then so is $WU$ with $U$
unitary. The change
\begin{equation} \label{ampli2}
W \longrightarrow W U
\end{equation}
is a {\em gauge transformation} with respect to a natural
gauge potential as we shall see later on. Here we need the
following: Let $W_j$ be an amplitude of $\rho_j$. By the help
of gauge transformations we can alter $W_1^{\dag}W_2$ to
$U_1^{\dag}W_1^{\dag} W_2 U_2$. Therefore, there are amplitudes
with $W_1^{\dag} W_2 \geq \0$.

A pair of amplitudes $W_j$ of $\rho_j$, $j=1,2$, is called
{\em parallel} if
\begin{equation} \label{ampli3}
 \0 \leq W_1^{\dag} W_2 = W_2^{\dag} W_1 \; .
\end{equation}
Parallel amplitudes allow to ``take the root'' in (\ref{expli1}).
Indeed,
\begin{displaymath}
(W_1^{\dag} W_2)^2 = (W_1^{\dag} W_2) (W_2^{\dag} W_1)
= W_1^{\dag} \rho_2 W_1 \; .
\end{displaymath}
By polar decomposing we can write $W_j = \sqrt{\rho_j} \, U_j$.
From (\ref{ampli3}) it follows that for any
pair $W_1$, $W_2$, of parallel amplitudes there are
unitaries $U_j$ such that
\begin{equation} \label{ampli4}
W_1^{\dag} W_2
= U_1^{\dag} (\rho_1^{1/2} \rho_2 \rho_1^{1/2})^{1/2} U_1
= U_2^{\dag} (\rho_2^{1/2} \rho_1 \rho_2^{1/2})^{1/2} U_2
\end{equation}

\subsection{A gauge invariant}
Let $W_1$ be invertible. The operator $W_2 W_1^{-1}$
remains invariant if $W_j \to W_j U$, see (\ref{ampli2}).
For invertible {\em parallel} amplitudes $W_j$ one gets
\begin{equation} \label{ampli5}
W_2 W_1^{-1} \geq \0 , \quad W_1 W_2^{-1} \geq \0
\end{equation}
The assertion can be seen from
\begin{displaymath}
W_2 W_1^{-1} =
(W_1^{-1})^{\dag} (W_1^{\dag} W_2) W_1^{-1}
 \geq \0 \; .
\end{displaymath}
By some algebraic manipulations one establishes
\begin{equation} \label{ampli6}
W_1 W_2^{-1} = \rho_1 \# \rho_2^{[-1]} \; , \quad
W_2 W_1^{-1} = \rho_2 \# \rho_1^{[-1]} \; .
\end{equation}
for invertible $\rho_j$. These operators play a role in the
no-broadcasting  theorem \cite{BCFJS96}. Another application
is in \cite{IB05}, \cite{AE07}.

\subsection{Amplitudes and Purification}
Let $|\varphi_1\rangle$, $|\varphi_2\rangle$ be purifying
vectors for $\rho_1$, $\rho_2$ in
$\cH \otimes \cH'$ with $\dim \cH \leq \cH'$. That means,
similar to (\ref{intro4}),
\begin{equation} \label{puramp1}
\T \, \rho_j A_j =
\langle \varphi_j, (A_j \otimes \1') \varphi_j \rangle
, \quad j = 1,2  \; .
\end{equation}
With two purifying vectors at hand we define $\nu_{12}$
to be the partial trace of
$|\varphi_2 \rangle\langle \varphi_1|$ over $\cH'$,
\begin{equation} \label{puramp2}
\nu_{12} := \T' |\varphi_2 \rangle\langle \varphi_1|
 , \quad
\T \, \nu_{12} A =
\langle \varphi_1| (A \otimes \1' |\varphi_2)
\end{equation}
for all operators $A$ acting on $\cH$.

Now let $W_1$, $W_2$ denote a pair of amplitudes for our two
states $\rho_1$, $\rho_2$. We choose a maximally entangled
vector $\varphi$ in $\cH \otimes \cH'$. Such a vector
purifies the maximally mixed state on $\cH$. It follows that
\begin{equation} \label{puramp3}
|\varphi_j \rangle = (W_j \otimes \1') |\varphi\rangle,
\quad
\nu_{12} = W_2 W_1^{\dag} \; .
\end{equation}
and $|\varphi_j$ purifies $\rho_j$ for $j = 1,2$.
\medskip

The Cauchy-Schwarz inequality bounds the right hand side of
(\ref{puramp2}) by
\begin{displaymath}
\langle \varphi_1 | A_1^{\dag} A_1 | \varphi_1\rangle \,
\langle \varphi_2 | A_2^{\dag} A_2 | \varphi_2\rangle =
(\T \, A_1^{\dag} A_1 \rho_1) \,
(\T \, A_2^{\dag} A_2 \rho_2)
\end{displaymath}
and, therefore, $\nu_{12}$ is restricted by
\begin{equation} \label{puramp5}
| \T \, \nu_{12} A_1^{\dag} A_2 |^2 \leq
(\T \, A_1^{\dag} A_1 \rho_1) \,
(\T \, A_2^{\dag} A_2 \rho_2)
\end{equation}
Now we assert
\begin{equation} \label{puramp6}
\Pr(\rho_1, \rho_2) = \sup  | \T \, \nu_{12} |^2
\end{equation}
where $\nu_{12}$ runs through all operators satisfying
(\ref{puramp5}).\\
The right hand side cannot be smaller neither
than $\Pr(.,.)$ as defined by (\ref{intro3}) nor than the
squared $F(\rho_1, \rho_2)$ as given by (\ref{expli1}).
To  see the other direction we choose parallel amplitudes
and set $A_1^{\dag} A_1 = W_2 W_1^{-1}$ and
$A_2^{\dag} A_2 = W_1 W_2^{-1}$ for invertible density
operators. Now, as a short calculation like
\begin{displaymath}
\T \, \rho_1 W_2 W_1^{-1} = \T \, \rho_2 W_1 W_2^{-1}
= F(\rho_1, \rho_2)
\end{displaymath}
shows: (\ref{puramp6}) can be saturated.
By continuity the degenerate cases can be settled.

The latter reasoning also shows
\begin{equation} \label{puramp7}
\Pr(\rho_1, \rho_2) = \inf_{A > \0}  (\T \, \rho_1 A) \,
 (\T \, \rho_2 A^{-1})
\end{equation}
because every term of the right side must be larger than
any $\T \, \nu_{12}$ again by Cauchy - Schwarz.

As an byproduct we see that equality in (\ref{puramp7})
can be reached with equal factors on the right. This can
be used to see the equivalence of (\ref{puramp7}) with
\begin{equation} \label{puramp9}
F(\rho_1, \rho_2) = \frac{1}{2} \inf_{A > \0}
\left( \T \, \rho_1 A \, + \,  \T \, \rho_2 A^{-1} \right)
\end{equation}

In full generality, i.~e.~for unital C$^*$-algebras,
(\ref{puramp7}) has been proved in \cite{inf2} using
an idea of \cite{inf1}.

\subsection{Concavity and monotonicity}
(\ref{puramp9}) is particulary suited to prove
\underline{concavity}. It presents fidelity by
an infimum of linear functions and tells us
\begin{equation} \label{puramp10}
F(\sum \lambda_j \rho_j, \sum \mu_k \rho_k) \geq
\sum \sqrt{\lambda_j \mu_j} \, F(\rho_j, \omega_j)
\end{equation}
for all choices of non-negative $\lambda_j, \mu_k$. Combined with
(\ref{expli1}) one concludes that equality holds in
(\ref{puramp10}) if and only if
\begin{equation} \label{puramp11}
\rho_j \omega_k = \0 \, \hbox{ for } \, j \neq k \; .
\end{equation}
Regarding concavity of $\Pr(.,.)$ see \cite{AlUh83}.

Let us prove \underline{monotonicity} as asserted in
subsection 2.1: Returning to (\ref{puramp9}),
let $\Phi$ be a trace preserving positive map and $\Psi$
the adjoint of $\Phi$. $\Psi$ is positive and unital.
Using Choi's inequality $\Psi(A^{-1}) \geq \Psi(A)^{-1}$,
valid for unital positive maps and positive $A$, one
proceeds according to
\begin{displaymath}
 \T \, \Phi(\rho_2) A^{-1} = \T \, \rho_2 \Psi(A^{-1})
\leq \T \, \rho_2 \Psi(A)^{-1}
\end{displaymath}
However, the set of positive operator of the form $\Phi(A)$ is
not larger than the set of all positive operators. Asking for
the minimum over all $A \geq \0$ we arrive at
\begin{equation} \label{puramp12}
F(\rho_1, \rho_2) \leq F(\Phi(\rho_1) , \Phi(\rho_2))
\end{equation}
for all stochastic, i.~e.~trace preserving positive maps.

(\ref{puramp12}) justifies the assertion (\ref{intro8}). Remark
in addition that we do not need complete positivity of $\Phi$.

\section{Geometric Phases}
At first let us remind some essentials of phases for
pure states. Starting with a curve of pure states
$\pi_s$, $0 \leq s \leq r$, one asks for resolutions
or lifts
\begin{equation} \label{phase1}
s \to |\psi_s\rangle, \quad \pi_s = |\psi_s \rangle\langle
\psi_s|, \quad 0 \leq s \leq r \; .
\end{equation}
Given an initial vector $|\psi_0\rangle$
the {\em parallel (or adiabatic) transport condition}
provides a unique lift of a given (regular enough) curve
$s \to \pi_s$. The condition is completely independent
of dynamics and reads
\begin{equation} \label{phase2}
 \langle \psi_s| \frac{d}{ds} |\psi_s\rangle =
\langle \psi_s| \frac{d}{ds} |\psi_s\rangle^* \; .
\end{equation}
(\ref{phase2}) is determining {\em geometric (Berry) phase}
of closed curves $s \to \pi_s$.

Any lift (\ref{phase1}) can be obtained
by a {\em gauge transformation}
\begin{equation} \label{phase3}
|\psi_s\rangle \mapsto |\chi_s\rangle =
\epsilon_s |\psi_s\rangle , \quad |\epsilon_s| = 1 \; .
\end{equation}
Let us abbreviate the derivatives $d/ds$ by a dot. One finds
\begin{equation} \label{phase4}
\langle \dot \chi | \dot \chi \rangle =
\langle \dot \psi | \dot \psi \rangle  +
|\dot \epsilon|^2 \; .
\end{equation}
Hence, a parallel lift comes with the {\em shortest Hilbert
length} within all lifts (\ref{phase1}) of a given curve
$s \to \pi_s$ of pure density operators. This minimal possible
length is the {\em Fubini-Study length} of $s \to \pi_s$.

\subsection{Parallelity and the minimal length condition}
At first we extend the minimal length condition.\\
Consider a given curve of density operators and their possible
amplitudes,
\begin{equation} \label{gphase1}
s \to \rho_s, \quad s \to W_s, \quad \rho_s = W_s W_s^*
\end{equation}
Required: Neighbored amplitudes should be ``approximately''
parallel in the understanding of (\ref{ampli5}). It results
the parallelity condition \cite{Uh86a}
\begin{equation} \label{gphase2}
(\frac{d}{ds} W_s)^{\dag} W_s -
W_s^{\dag} (\frac{d}{ds} W_s) = \0 \; .
\end{equation}
In the following we abbreviate $(d/ds)W$ by $\dot W$.
Apart from some singular cases one can go into (\ref{gphase2})
by an ansatz $\dot W = G W$. After inserting one finds
\begin{equation} \label{gphase3}
\dot W_s = G_s W_s, \quad  G_s^* = G_s \; .
\end{equation}
$G_s$ can be determined by differentiating $\rho = WW^{\dag}$
and inserting (\ref{gphase3}),
\begin{equation} \label{gphase4}
\dot \rho_s = \rho_s G_s + G_s \rho_s \; .
\end{equation}
$\dot \rho$ is a tangent at $\rho$. $G$ is a cotangent at $\rho$
with respect to the Riemann metric form $\T \, G^2 \rho$. Indeed,
the latter is the Riemann metric determined by the
{\em Bures distance.}
In particular,
\begin{equation} \label{gphase5}
\hbox{length}_{{\rm Bures}}[s \to \rho_s] =
 \int (\T \, G_s \rho_s G_s)^{1/2} ds  \; .
\end{equation}
By (\ref{gphase3}) we now see that
\begin{equation} \label{gphase6}
\hbox{length}_{{\rm Bures}}[s \to \rho_s] =
\int (\T \, \dot W_s \dot W_s^{\dag})^{1/2} ds \; ,
\end{equation}
provided $W_s$ satisfies the parallelity condition
(\ref{gphase2}). One can show that the latter integral cannot
become smaller by any gauging $W_s \to W_sU_s$ of the parallel
amplitudes $W_s$.

The Bures distance \cite{Bu69} is well described in \cite{BZ06}
and in \cite{CrUh}, \cite{NieChubook}. That it is a distance of
a Riemann metric has been seemingly overlooked for long,
\cite{Uh92b}, \cite{BC94}. A systematic way to find the metric
tensor is in \cite{Di95}. $\dim \cH =2$ is discussed in
\cite{BZ06}, $\dim \cH = 3$ in \cite{Sl01}. See \cite{BR97}
how to solve (\ref{gphase4}).

\section{A gauge theory}
We look for a gauge potential (connection form) $\mathbf{A}$
with the following property:
The restriction $A_s ds$ of $\mathbf{A}$ to $s \to W_s$
should vanish if and only if $W_s$ satisfies the parallelity
condition (\ref{gphase2}). Being a gauge potential we
have to have firstly
\begin{equation} \label{gauge1}
\bA + \bA^{\dag} = \0 \; ,
\end{equation}
and, secondly, a gauge transformation $W_s \to W_s U_s$ must result in
\begin{equation} \label{gauge2}
\bA  \to U^{-1} \bA  U + U^{-1} \rd U \; .
\end{equation}
Let us find $\bA$ for invertible\footnote{If the rank changes,
the problem becomes sophisticated.} $W_s$. We rewrite
(\ref{gphase4}) as a relation between operator valued
differential 1-forms,
\begin{equation} \label{gauge3}
\rd \rho = \bG \rho + \rho \bG , \quad \bG^{\dag} = \bG \; .
\end{equation}
We now define $\bA$ by
\begin{equation} \label{gauge4}
\rd W = \bG W + W \bA \; .
\end{equation}
By the help of (\ref{gauge3}) and (\ref{gphase1})
one can establish (\ref{gauge1}) and (\ref{gauge2}). Thus,
(\ref{gauge4}) indeed defines a gauge potential.
To see that it fulfills our requirement one inserts (\ref{gauge4})
into (\ref{gphase2}), \cite{DG90}. The result is
\begin{equation} \label{gauge5}
W^{\dag} \rd W - (\rd W^{\dag}) W =
W^{\dag} W \, \bA + \bA \, W^{\dag} W
\end{equation}
proving that the parallel condition can be implemented by a
genuine gauge theory, \cite{Uh91a}. To identify $W^{\dag}W$
let us shortly return to the purification process, attaching
$|W\rangle = (W \otimes \1') |\varphi\rangle$ to a maximally
entangled $|\varphi\rangle$ and an amplitude $W$. Taking the
partial trace over $\cH$ in $\cH \otimes \cH'$ one obtains
the density matrix $W^{\dag} W$ belonging to $\cH'$.

As we have seen, the phase transport along curves of density
operators can be described either by a minimal length condition
or by a gauge theory. This suggest further relations between
the Bures Riemann metric and the gauge potential. One of
them concerns the curvature form $\rd \bA + \bA \wedge \bA$
and the Cartan curvature form of the metric. Remarkable enough
it turns out \cite{Uh91a} that
\begin{equation} \label{gauge6}
(\rd \bG - \bG \wedge \bG) W +
W \, (\rd \bA + \bA \wedge \bA) = 0 \, .
\end{equation}

\cite{ChJ04} is a general reference to the geometric phase for
general states. For relations to Einstein-Yang-Mills systems see
\cite{To93}. For comparison with other approach see \cite{Sl02}.
A treatment of the $\dim \cH = \infty$ case is
in \cite{CM01}. Other aspects, including the problem of
experimental verifications are in \cite{AKSO07} and
in \cite{Ku07}, where further references can be found.

\section{Conclusion}
The paper describes a small but nevertheless rich
part of what may be called the ``non-dynamical basis''
or the ``grammar'' of quantum physics. By the rising of
quantum information theory its importance has become  much
more evident then before, though it has been clearly seen
already in the so-called algebraic approach to quantum field
theory and statistical physics. Of course, experimental progress
can be made only in combination with dynamics, concrete Hamiltonians
and so on. On the other hand, the rules, we have had addressed
in the paper, are of such a generality that one
can scarcely believe they can be derived or proved from
specially chosen dynamics. To the belief of the author the
things are just opposite: These general rules are setting conditions
for possible forms of dynamics, including space and time.
\medskip

\underline{Acknowledgement}. I thank B.~Crell and P.~M.~Alberti
for support and help.

\end{document}